\documentclass[aps,prb,twocolumn,groupedaddress,superscriptaddress,showpacs,floatfix]{revtex4}

\usepackage[dvips]{graphicx}
\usepackage{xcolor}

\begin{document}

\title{Interfaces between buckling phases in Silicene: \textit{Ab initio}
density functional theory calculations }

\author{Matheus P. Lima}
\email[]{mplima@if.usp.br}
\affiliation{Instituto de F\'isica, Universidade de S\~ao Paulo, CP 66318, 05315-970, S\~ao Paulo, SP, Brazil.}

\author{A. Fazzio}
\email[]{fazzio@if.usp.br}
\affiliation{Instituto de F\'isica, Universidade de S\~ao Paulo, CP 66318, 05315-970, S\~ao Paulo, SP, Brazil.}

\author{Ant\^onio J. R. da Silva}
\email[]{jose.roque@lnls.br}
\affiliation{Instituto de F\'isica, Universidade de S\~ao Paulo, CP 66318, 05315-970, S\~ao Paulo, SP, Brazil.}
\affiliation{Laborat\'orio Nacional de Luz S\'{\i}ncrotron, CP 6192, 13083-970, Campinas, SP, Brazil.}

\begin{abstract}
The buckled structure of silicene leads to the possibility of new kinds of line 
defects that separate regions with reversed buckled phases. In the present work 
we show that these new grain boundaries have very low formation energies, one 
order of magnitude smaller than grain boundaries in graphene. 
These defects are stable along different orientations, and they can all 
be differentiated by STM images. All these defects present local dimerization
between the Si atoms, with the formation of $\pi$-bonds. As a result, these
defects are preferential adsorption sites when compared to the pristine region. 
Thus, the combination of low formation energy and higher reactivity of these defects 
may be cleverly used to design new nano-structures embedded in silicene.
\end{abstract}
\pacs{61.72.-y,61.46.-w,31.15.E-}

\maketitle

\section{Introduction}

The first in-lab observation of silicene\cite{silicene_first} has attracted the attention
of the community that studies new materials due to their similarities with graphene\cite{gr}.
For this reason, efforts have been employed to understand and control the growth of silicene\cite{exp1,exp2,exp3,exp4,exp5,exp6}.
These experimental measurements confirmed some theoretical 
predictions done several years before \cite{silicene_first_dft1,silicene_first_dft2}. 
Recent works have suggested that silicene is a potential candidate for applications in nano-electronics\cite{review_on_silicene}. 
However, having in mind device development and integration, it is fundamental to have a deep understanding
of the influences that defects, doping, interactions with substrates, external fields and magnetic moments have on the properties of this material in order to fully exploit
the possibilities of silicene in nanotechnology. Currently, these tasks represent great challenges and 
the necessary knowledge is still in its infancy.

When compared to graphene, silicene has a new and important ingredient: the presence of a buckled structure.
The understanding of the effects caused by this extra feature is crucial to explore the full potential of this material.
Few consequences of this buckling pattern have already been investigated; among them we can mention:
i) the increase of spin-orbit coupling, enhancing the Quantum Spin Hall Effect\cite{so};
ii) the possibility to tune the energy gap and the topological phase by the application of an external electric field perpendicular 
to the sheet\cite{efield1,efield2,efield3}; and iii) increase of the surface reactivity\cite{add-metals}.

Due to the buckling, there are, in fact, two equivalent, energy degenerate geometric phases ($\alpha$ and $\beta$) in silicene.
The $\alpha$ phase has a given atom shifted up and its first neighbors shifted down, 
whereas in the $\beta$ phase the shifts are reversed. 
A new possibility is thus the co-existence of both the $\alpha$ and $\beta$ phases in the same sample.
This situation may be created by different causes, such as a peculiar growth mechanism or the interaction with the substrate, since such interactions 
may pin down particular phases in different regions of the sample.

\begin{figure}[h!]
\begin{center}
\includegraphics[width=8.5cm]{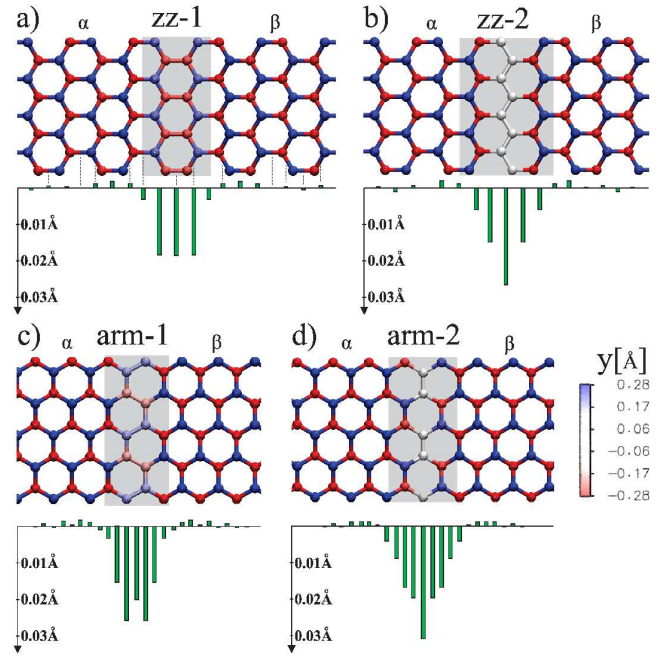}
\caption{\label{fig-geom} (Color Online) Fully relaxed geometries for the (a) zz-1 interface in the zigzag direction, (b) zz-2 interface
in the zigzag direction, (c) arm-1 interface in the armchair direction, (d) arm-2 interface in the armchair direction.
The atoms are colored according to their out-of-plane dislocation ($y$). The gray rectangles emphasize the interface region.
Below each structure is presented the bond length deviations from the pristine silicene ($d^{bulk}-d$).}
\end{center}
\end{figure}
In this article we investigate these new kinds of line defects created at the 
interface between the $\alpha$ and $\beta$ phases in silicene. These buckling phase interfaces 
may occur along different orientations (zigzag, armchair or intermediated chiral geometries), 
and have a much lower energy than grain boundaries in graphene.
The main characteristic of such defects is the presence of $\pi$ bonds
between the interface atoms caused by an out-of-plane dislocation rearrangement. 
For the zigzag direction, the formation of $\pi$ bonds is more clear, leading
to a slightly lower formation energy when compared to any other directions.
We analyse in detail how these interfaces could be observed and differentiated in Scanning Tunneling Microscopy (STM) images. 
Furthermore, we show that the reactivity is enhanced at these interface regions, exemplified by demonstrating that 
$Au$ atoms have a lower binding energy when adsorbed over these buckling phase interfaces in comparison with the pristine regions. 

\section{Computational details}

Our results were obtained with {\it ab-initio} density functional theory calculations.
We used the SIESTA code\cite{siesta} within the Local Density Approximation\cite{lda} 
(LDA) for the exchange correlation functional.
Two dimensional ($2D$) periodic boundary conditions were employed in all calculations, with
a grid of $50\times 50$ k-points in the unit cell,  a Double-$\zeta$ polarized basis, and a mesh cut-off of $300~Ry$ to define the grid in the real space. 
A vacuum of 20~\AA~ is sufficient to avoid undesirable interactions between the
periodic images of silicene sheets.
For the pristine system we obtain a lattice constant of $3.854$~\AA, bond lengths of $2.281~$ \AA, 
and out-of-plane dislocations of $\pm 0.25$~\AA.
These parameters are consistent with earlier results found in the literature.\cite{silicene_first_dft1,theo1,theo2}.
We also calculated a barrier of $35~meV/$atom to revert the buckling phase in the whole system. 

The simulations of linear defects are done with two complementary interfaces to reach the 2D periodicity in the 
sheet plane. 
To generate fully relaxed geometries we first take
a supercell with the pristine geometry. Subsequently, we revert the out-of-plane
dislocation only in a certain region where we wish to observe the inverted phase. Subsequently, we perform 
a fully conjugate gradient relaxation using a force criterion of $0.015~eV/$\AA~
to quench the forces in the whole system. As expected, the major modifications will occur only close to the phase interface.

\section{Results}
\subsection{fully relaxed geometries}

The fully relaxed geometries (which remain planar) for the interfaces considered in this work 
are shown in Fig. \ref{fig-geom}.\cite{chiral-com}
We considered four kinds of interfaces, being two along the zigzag direction, and two along the armchair
direction. 
In (a) we show the zz-1 interface along the zigzag direction, which has pairs of neighboring atoms with the same out-of-plane dislocation (y).
In (b) we show the zz-2 interface along the zigzag direction. In this case, there is a line of zigzagged atoms with $y=0$.
In (c) it is depicted the arm-1 interface, which occurs along the armchair direction. Here, the interface atoms form a line with dimerized 
out-of-plane dislocations, which means two atoms with positive out-of-plane dislocations 
followed by two atoms with negative out-of-plane dislocations. And, in (d) it is presented the 
arm-2 interface along the armchair direction. There are pairs of neighboring atoms with $y=0$ at this interface.
Below each structure we present the bond length variation from the pristine distance. Although the 
interfacial atoms have a bond length around $0.03$\AA~ shorter than the pristine atoms, the 
interatomic distances are quickly reestablished as one moves away from the line defect. Such analysis 
allows to conclude that the width of 
these line defects is not more than $7$\AA.

\subsection{\label{sec-EE} Electronic Structure}

\begin{figure}[t!]
\begin{center}
\includegraphics[width=8.5cm]{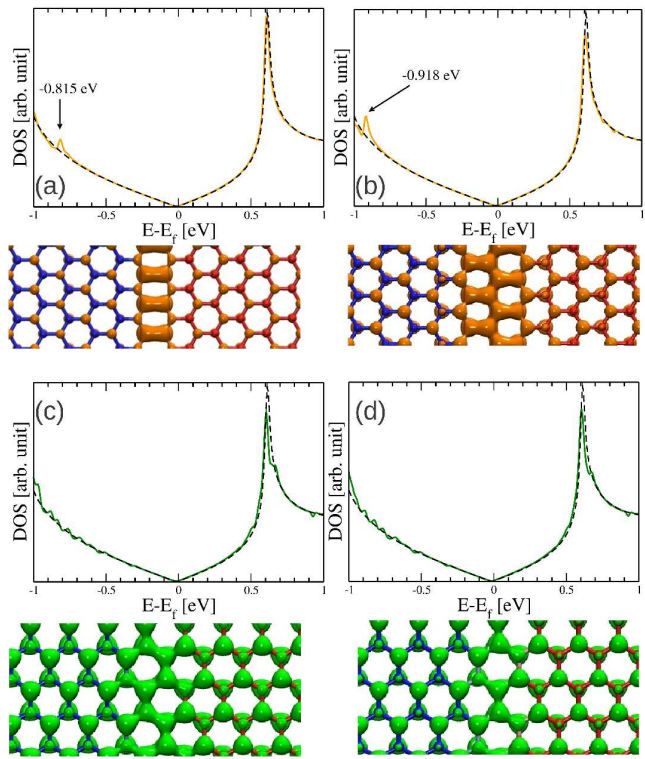}
\caption{\label{fig-DOS} (Color Online) DOS and LDOS for the $\alpha$-$\beta$ interfaces: (a) zz-1 (b) zz-2, (c) arm-1 and (d) arm-2.
                      For (a) and (b) the LDOS were calculated around the peaks localized at $-0.814~eV$ and $-0.918~eV$
                      below the Fermi energy, respectively. An energy window of $\pm0.025~eV$ was adopted. 
                      For (c) and (d) the LDOS were calculated integrating the states within the energy range $-0.9~eV \leq E-E_f \leq -0.4~eV$.}
\end{center}
\end{figure}

Aiming to understand the influences of such line defects in the electronic structure of silicene, we calculated the
Density of States (DOS) within an energy window of $\pm 1~eV$ around the Fermi energy. 
We present the results in Fig. \ref{fig-DOS}. 
For the zz-1 interface, shown in (a), the DOS is very similar to the pristine case, with 
exception of a sharp peak localized at $-0.815$ below the Fermi energy ($E_f$). The Local Density of States (LDOS)
calculated within an energy window of $\pm 0.025~eV$ centered at this peak is shown below its DOS.
This LDOS reveals that this peak is primarily composed by states with a $\pi$-bond signature localized 
at the interface atoms.
The narrow shape of this resonance peak indicates that the zz-1 defect levels 
have a small dispersion as well as a small coupling to the 
rest of the system. In Fig. \ref{fig-DOS} (b) we present the results for the zz-2 interface. 
Again, only a sharp peak is present in the DOS, however, its energy is lower ($-0.918~eV$), and 
its amplitude is higher. Similarly to the zz-1 case, the LDOS centered at this peak allows us to conclude 
that these states have a $\pi$ bond signature, and a small coupling to the rest of the system.  
Therefore, the creation of both kinds of zigzag defects leads to the formation of $\pi$ states localized 
at the interface atoms.

To better illuminate the understanding about the influence of such 
linear defects in the electronic structure of silicene, we modeled 
the system with an effective first nearest-neighbour tight-binding (TB) Hamiltonian, given by: 
\begin{eqnarray}
H=\sum_{<i,j>}-tc^\dagger_ic_j+\sum_i U_i c^\dagger_ic_i+h.c.
\end{eqnarray}
$c^\dagger_i$ ($c_i$) creates (annihilates) an electron at the site $i$, $t$ is the 
transfer integral, and $U_i$ is the on-site energy. 

\begin{figure}[t!]
\begin{center}
\includegraphics[width=8.5cm]{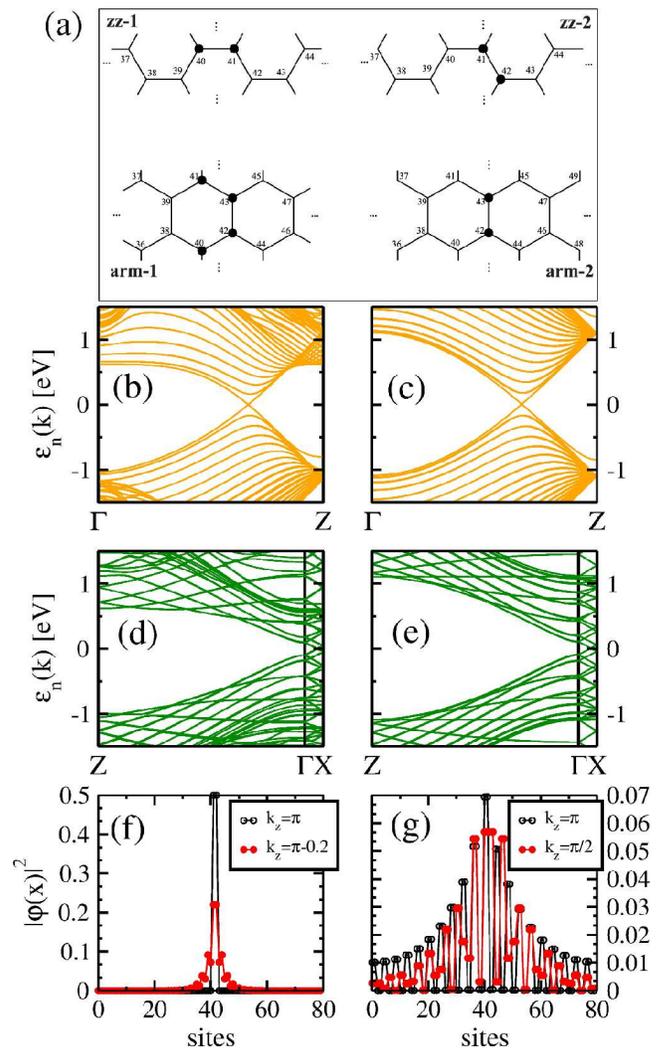}
\caption{\label{fig-tb} 
(a) Black circles highlight the re-hybridized sites where $U_{i}$ must be modified
for the zz-1, zz-2, arm-1 and arm-2 defects.
Energy bands for zz-1 structure calculated with (a) DFT and 
(b) our tight-binding model; (c) and (d) show similar calculations for the arm-1 structure. Tight-binding wave-functions
for the highest-occupied energy band ($k_x=0$) for the (e) zz-1; and (f) arm-1 structures. Here, the tight-binding calculations 
consider a total of 80 sites ($N=80$).}
\end{center}
\end{figure}

Considering this model, the main effect caused by the buckling phase inversion 
are changes in the on-site energy $U_{i}$ only for the interfacial atoms 
due to the local re-hybridization occurring at them\cite{hopp-com}. 
However, note that the disposition of these interfacial atoms is distinct for each structure. 
In fig \ref{fig-tb} (a) we mark
with black circles the re-hybridized sites where $U_{site}$ must be modified.
We adjust the tigh-binding parameters to better fit with the energy bands calculated with DFT. 

In fig \ref{fig-tb} (b) and (c) we compare the DFT and TB energy bands for the zz-1 interface, respectivelly. 
Despite some discrepancies far from $E_f$, the general behavior is essentially the same for both DFT and TB calculations.
For this defect, the adjusted parameters are $t=1.1eV$, and $U_{i}=0.25t$ ($U_i=0$ for bulk atoms). 
The positive value of $U_{i}$ brings the defect states to the top of the occupied energy bands. 
Analyzing the Hamiltonian (see appendix \ref{app1}), it is possible to note that 
the effective coupling of the defect atoms highlighted in Fig. \ref{fig-tb} (a) to the rest of the system
have a form of $-t(1+e^{\pm ik_z})$, and are thus negligible close to $k_z=\pi$.
The modulus square of the TB wave function at $k_z=\pi$
is depicted in Fig. \ref{fig-tb} (f), and can be clearly seen a strictly localized wave function. 
For other values of $k_z$, the states progressively become more dispersive as a consequence 
of the increasing of the effective coupling of the defect sites to the rest of the system.
The wave function for $k_z=\pi-0.2$ is also depicted in Fig. \ref{fig-tb} (f),
and a non-null contribution of the neighbouring sites induced by the increase of the effective coupling term is present.

We also compare, in Fig. \ref{fig-tb} (d) and (e), the DFT and TB results for the 
arm-1 defect, respectively. Here, the adjusted parameters are $t=1.1eV$, and $U_{i}=0.15t$.
Again, there is a great overall agreement between the results calculated with both methods, 
and due to the positive value of $U_{i}$, the defect states once again lie at the top of the highest occupied band. 
However, contrary to the zigzag direction, in the armchair direction the effective coupling between the defect sites and 
the rest of the system never vanishes for any $\vec{k}$ point. 
We illustrate this behavior in Fig. \ref{fig-tb} (g), where we show the 
modulus square of the wave function for $k_z=\pi$ and $k_z=\pi/2$. It can be clearly seen that, even though
the states are more localized at the defect sites, 
a contribution at sites of distant neighbours are present. 
The numerical TB results also indicate that the states at the central region 
of the valence band (around $k=\pi/2$) are slightly more localized than the 
states close to the $\Gamma$  and $Z$.

Now, let us focus again in the results obtained with calculations based on DFT.
The DOS are presented in Fig. \ref{fig-DOS} (c) and (d) for the arm-1 and arm-2 interfaces,
respectively. The behavior is quite similar for both armchair interfaces:
i) there are no peaks associated with any localized states; ii) The DOS are very similar 
to the pristine case, except for small oscillations caused by the 
interaction between the line defects and its complementary images (see appendix
\ref{app2}).
Since our TB model indicates that the defect states are mainly localized at the central part of the valence band, 
we calculated the LDOS for an energy range from $E-E_f=-0.9$ to $-0.4~eV$, for both arm-1 and arm-2 cases. 
In both cases, it is possible to see a slightly greater contribution of the defect sites,
even considering that the defect levels are weakly localized, and some bulk states are been included in this LDOS.

An important remark to be made is that, even though these interfaces create 
defect states at the top of the highest occupied energy band, close to $E_f$
they are always delocalized. As a result, 
the modifications in the DOS very close to the Fermi level in all cases are negligible.
For example, the Dirac Cone characteristic V-shape is fully maintained.  Even with the inclusion of the spin-orbit coupling in the 
calculations, the differences between the pristine and the defective systems are negligible around $E_f$.
Therefore, these defects will be quasi-invisible to measurements that depend solely on the Fermi-surface.

\subsection{Simulations of scanning tunneling microscope images}
  
\begin{figure}[t!]
\begin{center}
\includegraphics[width=8.5cm]{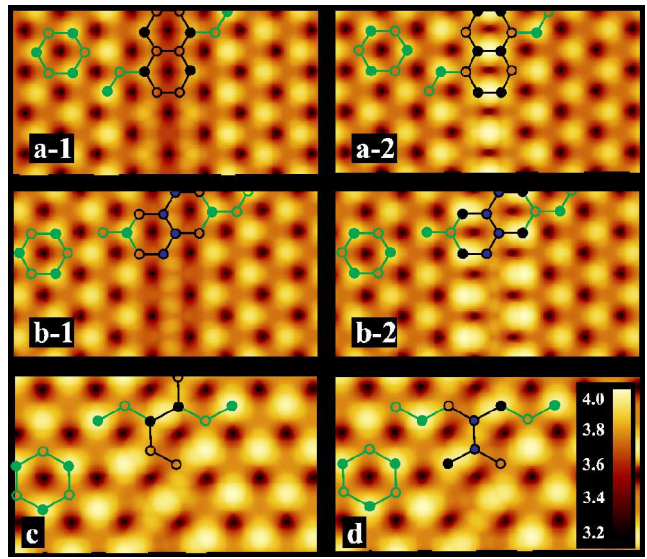}
\caption{\label{fig-stm} (Color Online) Images of the STM simulations for the buckling phase interfaces. 
(a-1) and (a-2) show the zz-1 interface; (b-1) and (b-2) show the zz-2 interface;
(c) shows the arm-1 interface; and (d) shows the arm-2 interface.
The filled (non-filled) circles represent Si atoms with up (down) shifts. The lines are guides to the eye. The scale bar
represent the tip height in \AA.}
\end{center}
\end{figure}

We show in Fig. \ref{fig-stm} that it is possible to differentiate the various interfaces with scanning tunneling microscopy (STM)
images. We used the Tersoff-Ramann theory\cite{tersoff.ramann}, with a bias voltage of $1.0~V$, considering 
occupied states and the constant current model. 
STM images were commonly used to experimentally investigate silicene\cite{silicene_first,exp1,exp2,exp3,exp4,exp5,exp6}.
Similarly to the experiments, our simulations show 
a clear signature of the buckling pattern at the pristine region: the signal presents high spots over the up-shifted atoms, 
low spots over the down shifted atoms, and absence of signal at the center of the hexagons\cite{exp-buckled}.
For the buckling interfaces, the out-of-plane displacements create distinct patterns depending on the interface symmetry, as shown in detail in Fig. \ref{fig-stm}.
In (a-1) we present the STM image for the zz-1 interface. The position of the interface atoms are indicated by black circles.
In this picture, there is a clear line of low signals along the dimerised central interface atoms because they 
are down-shifted with respect to the STM tip. If we consider the other possible orientation of the STM tip, these same atoms would now be up-shifted, and we would obtain a line of high signals, as shown in Fig. \ref{fig-stm} (a-2). An interesting feature of this image is the clear signal of $\pi$ bonds between the central atoms of the defect.
For the zz-2 interface there are also two possible images, depending on the tip orientation. These images are shown in Fig. \ref{fig-stm} (b-1) and (b-2). 
Here, the source of the differences is the fact that all first-neighbour atoms to the interface atoms have the same out-of-plane dislocation. Well defined $\pi$ bonds can also be 
identified between interface atoms and its first neighbours via STM images.

For the interfaces in the armchair direction, differently from those in the zigzag direction, the STM images do not depend on the tip orientation, as shown in Fig. \ref{fig-stm} (c) and (d)
for the arm-1 and arm-2 interfaces, respectively. In both cases, the out-of-plane rearrangement can be 
clearly inferred from STM images, and the $\pi$ bonds can also be detected. Thus, we have shown that it is experimentally possible to use STM images to identify and differentiate all these distinct interfaces.

\begin{table}
\caption{\label{tab1} Formation Energy per unity length ($U^{form}$ in meV/$\AA$) for the 
line defects considered in this work.}
\begin{ruledtabular}
\begin{tabular}{cccccc} 
          & zz-1 & zz-2 & arm-1 & arm-2 & $\theta=19.11^\circ$\\
\hline
$U^{form}$  & 17.5   & 17.9  & 20.9   & 20.3 & 22.8 \\  
\end{tabular}
\end{ruledtabular}
\end{table}

\subsection{Formation energies}
\begin{figure}[h!]
\begin{center}
\includegraphics[width=8.5cm]{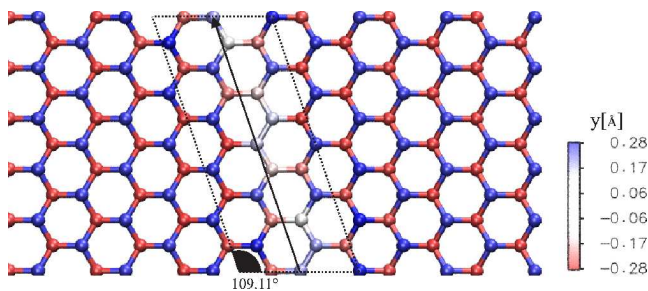}
\caption{\label{fig-chiral} (Color Online) Geometry for a chiral line defect. The angle between the
defect direction and the pure zigzag direction is $19.11^\circ$. Here, the atoms are coloured
according to its out-of-plane dislocation. The arrow represent the lattice vector along the
defect direction.}
\end{center}
\end{figure}

The formation energies per unity length ($U^{form}$) for all the defects are presented in table \ref{tab1}. 
The interfaces along the zigzag direction have the lowest formation energies, because the $\pi$ bonds are more clearly formed, leading to a
slightly larger energy gain. Otherwise, the two zigzag and the two armchair interfaces have 
formation energies quite similar among themselves. Besides the zigzag and armchair directions, we also consider one intermediate case, with an angle $\theta=19.11^\circ$ with the zigzag direction. 
The fully relaxed geometry is shown in Fig. \ref{fig-chiral}, where can be seen
that  the rearrangement of the out-of-plane dislocations are more complex, leading to a 
higher formation energy due to the increase of the elastic contribution. We expect 
the same behavior for other chiral angles.

The values of $U_{form}$ presented in table \ref{tab1} indicate that the defects in the zigzag direction
will be more easily created at low temperatures.
Indeed, a recent work by Lan Chen {\it et al.} reports a spontaneous symmetry breaking phase transition in 
silicene over the Ag(111) substrate\cite{PRL110.085504}. They have shown, at low temperatures, a spontaneous 
creation of a 2D grid where the domains are triangles with alternating buckling phases separated by interfaces in the zigzag direction. 
Although the authors did not discuss the structural details of the triangles borders, it can be inferred from their STM images that they are the zz-2
interfaces described above. It is important to note that the particular triangular structure observed experimentally will most likely depend on the interaction between the silicene and the Ag substrate. This indicates that on the one hand the defects that we propose are very likely to be observed, but on the other hand the particular configuration that they will have on the silicene sheet will depend on experimental details such as substrate configuration and growth temperature.

It is important to stress that the values presented in table \ref{tab1} are one order of magnitude lower 
when compared to either grain boundaries or dislocation linear defects in graphene\cite{gb1,gb2}.
This is a consequence of to fact that grain boundaries in graphene always involve a chemical bond 
reconstruction, and the hexagons are converted in other polygons, like pentagons, heptagons, etc. 
This kind of reconstruction highly increases the formation energy of the linear defects in graphene.
On the other hand, as reported in this paper, the linear defects in silicene may be generated by 
another mechanism: an out-of-plane dislocation rearrangement, leading to lower values of 
the formation energy and, thus, a defect more likely to be created.

\subsection{Adsorption of Gold atoms}

Another interesting point related to these line defects is that they might be preferential adsorption sites. To investigate this question, we performed proof-of-concept simulations with a single gold atom adsorbed over the system.
In these simulations we used surpercells in which the Au atoms are separated by more than 18~\AA~ 
from its periodic image. This is necessary to sufficiently decrease the interaction between the Au atom and its images.
We choose Au atoms because there are previous experiments of single atom adsorption over the Si(100) 
surface\cite{si.surface}, which bear some similarities with the line defects we propose due to the presence of dimers. 
Furthermore, gold is frequently used in nano-devices.

\begin{figure}[!t]
\begin{center}
\includegraphics[width=8.5cm]{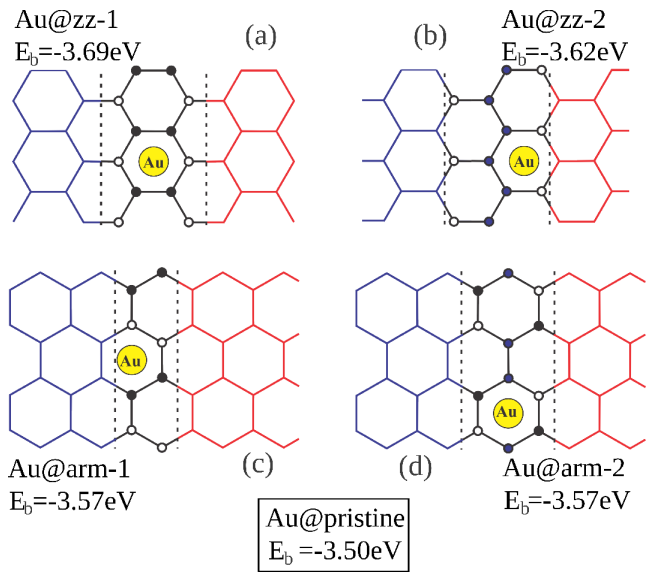}
\caption{\label{fig-ouro} (Color Online) Schematic representation of the most stable adsorption site of $Au$ over 
the (a) zz-1, (b) zz-2, (c) arm-1, and (d) arm-2 interfaces.}
\end{center}
\end{figure}

Considering a single gold atom over the pristine silicene, we found a binding energy ($E_b$)
of -3.50~eV when adsorbed at the hollow site. 
This adsorption site is energetically favourable by $0.38~eV$ and $0.66~eV$
when compared to the top and bridge configurations, respectively. Corroborating previous works, this binding 
energy indicates a much stronger bond in silicene when compared to graphene\cite{add-metals,tm-matheus,amft,miwa1,miwa2}.
To investigate the adsorption of Au over the buckling interfaces we considered several adsorption sites, 
and the energetically most favourable ones for each one of the four interfaces are shown in Fig. \ref{fig-ouro}.
In the same figure we also present $E_b$ for each one of these configurations. 
Confirming our expectations, the results show that all these defects are more reactive than the pristine region, indicating that they will be preferential absorption sites. This opens up a variety of possibilities for nano-engineering, where via adsorption of particular atoms or molecules at these sites on can taylor the properties of 1D channels embedded on the silicene sheet.

\section{Conclusions}
In conclusion, we have shown that due to the buckling structure of silicene, it is possible to have a new kind of low energy grain boundary associated with the reversal of the buckling phase, contrary to graphene, where the grain boundary defects have always high energy. In particular, we have investigated in detail these buckling phase interfaces, showing that:
i) modifications in the Density of States will appear far from the Fermi energy;
ii) The formation energy of these interfaces are very low, of the order of $(k_B T_{300K})$/\AA. These values are approximately one order of 
magnitude lower than grain boundaries in graphene;
iii) These structures can be experimentally identified by STM images;
iv) These interfaces are preferential adsorption sites when compared to the pristine region.
Therefore, these interfaces are important and most likely common defects in silicene, and may be thus used to control
the adsorption of atoms and molecules, which could lead to many possibilities in molecular engineering.

\section{Acknowledgements}
We would like to thank the Conselho Nacional de Desenvolvimento Cient\'ifico e
Tecnol\'ogico/Institutos Nacionais de Ci\^encia e Tecnologia do Brasil 
(CNPq/INCT), the Coordena{\c c}\~ao de Aperfei{\c c}oamento de Pessoal de N\'ivel 
Superior (CAPES), and the Funda{\c c}\~ao de Amparo \`a Pesquisa do Estado de S\~ao
Paulo (FAPESP).

\appendix

\section{\label{app1} Hamiltonian matrix elements for the tight-binding model of the
buckling phase interfaces.}

In this appendix we explicitly show the matrix elements for the effective model
used to study
the buckling phase inversion in silicene. We first construct
the Hamiltonian for the pristine silicene with a large supercell, and we then
simulate the
buckling inversion line defects by changing the on-site energy only for specific
sites.
\begin{figure}[b!]
\begin{center}
\includegraphics[width=8.5cm]{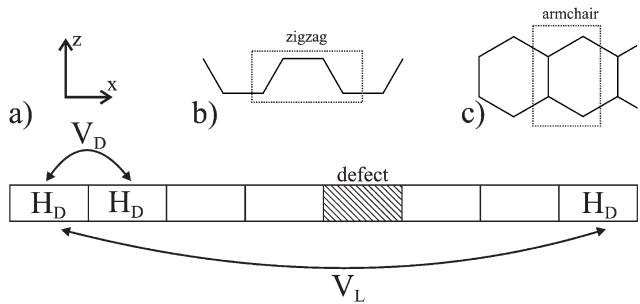}
\caption{\label{fig1-app1} In (a) we present a diagrammatic representation of the
Hamiltonian coupling terms.
In (b) and (c) the building blocks for the zigzag and armchair structures are
enclosed with dashed lines.
}
\end{center}
\end{figure}

For the pristine silicene, we construct a simple nearest neighbour tight-binding
Hamiltonian with
one effective orbital per site, with rectangular supercells having the
smallest possible size along the defect direction ($\hat{\bf{z}}$), and an
arbitrary size along the lateral
direction ($\hat{\bf{x}}$).
For both zigzag and armchair directions we used building blocks with 4 sites, as
shown in Fig. \ref{fig1-app1}
(b) and (c), respectively. Thus, the total pristine Hamiltonian can be written
as:
\begin{eqnarray}\label{genH}
H(k_x,k_y)=\left(\begin{array}{cccccc}
H_D         & V_D & 0 & 0  & \cdots & V_L \\
V_D^\dagger & H_D & V_D & 0 &\cdots & 0  \\
0 & V_D^\dagger & H_D & V_D &       & 0  \\
\vdots   &   &   & \ddots  &  &  \\
         &   &   &         &  &  \\
V_L^\dagger  & 0 & \cdots  & 0 & V_D^\dagger &  H_D  \\
\end{array}\right).
\end{eqnarray}

Here, $H_D$ is the building block Hamiltonian, $V_D$ is the coupling between
neighbouring building blocks,
and $V_L$ is the coupling between neighbouring supercells. A diagrammatic
representation of these interactions
is shown in Fig. \ref{fig1-app1} (a). Since the building blocks have 4 sites, all
these are $4\times 4$ matrices.
Particularly,  for the armchair direction these terms are given by:
\begin{eqnarray} 
H_D=-t\left(\begin{array}{cccc}
0             &       e^{-ik_z}   &      1     &     0    \\
e^{ik_z}    & 0                 &     0      &     1    \\
 1            & 0                 &     0      &     1    \\
0             &  1                &     1      &     0    \\
\end{array}\right);~~ \\ 
V_D=-t\left(\begin{array}{cccc}
0             &       0           &      0     &     0    \\
0           & 0                 &     0      &     0    \\
 1            & 0                 &     0      &     0    \\
0             &  1                &     0      &     0    \\
\end{array}\right);~~ \\ 
V_L=-t\left(\begin{array}{cccc}
0             &       0           &  e^{-ik_x} &     0    \\
0           & 0                 &     0      &     e^{ik_x}    \\
 0            & 0                 &     0      &     0    \\
0             &  0                &     0      &     0    \\
\end{array}\right). 
\end{eqnarray}
Where, $t$ is the transfer integral, $k_z$ ($k_x$) is the crystalline momentum
along (perpendicular to) the defect direction. For the zigzag direction, the
Hamiltonian
building block terms are written as:
\begin{eqnarray}\label{zz}
H_D=-t\left(\begin{array}{cccc}
0      &  c_k    &     0    &     0        \\
c_k^*  &  0      &     1    &     0        \\
0      &  1      &     0    &     c_k^*    \\
0      &  0      &     c_k  &     0        \\
\end{array}\right);~~ \\ 
V_D=-t\left(\begin{array}{cccc}
0    &  0   &    0    &     0    \\
0    &  0   &    0    &     0    \\
0    &  0   &    0    &     0    \\
1    &  0   &    0    &     0    \\
\end{array}\right);~~ \\ 
V_L=-t\left(\begin{array}{cccc}
0    &    0    &    0    &    e^{-ik_x}  \\
0    &    0    &    0    &    0          \\
0    &    0    &    0    &    0          \\
0    &    0    &    0    &    0          \\
\end{array}\right). 
\end{eqnarray}
Here, $c_k=\left(1+e^{ik_z}\right)$ is an effective coupling term. Note that
$c_k\rightarrow0$ when $k_z\rightarrow\pi$, as discussed in the
section \ref{sec-EE}.

The re-hybridization of the defect sites leads to changes in their on-site energy. 
Therefore, to simulate a single line defect it is necessary
to change the on-site energies (considered zero for the pristine sites) for a
single building block, as schematically shown in Fig. \ref{fig1-app1} (a).

\section{\label{app2} Oscillations in the Density of States of defects in the
armchair direction }

\begin{figure}[ht!]
\begin{center}
\includegraphics[width=8.5cm]{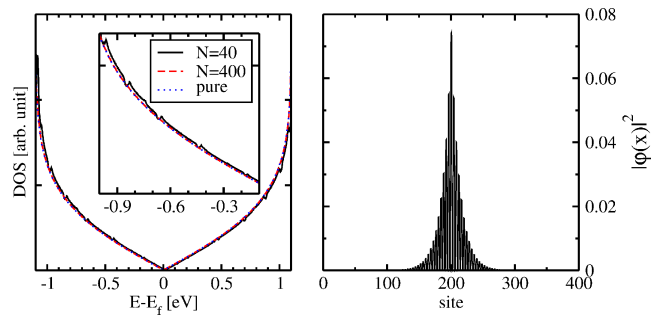}
\caption{\label{fig2} (Color Online)  (a) Density of  States (DOS) of silicene with the arm-2
defect varying the
distance between the line defect and its lateral image. For N=40 the distance is
$3.8nm$,
whereas for N=400 the distance is $38nm$. We normalize the absolute value or the
DOS by the respective number of
sites of the system. The inset shows a zoom of the DOS.
Also considering the arm-2 defect we show in (b) the modulus square of the wave
function of the valence band at $k_x=0$, and $k_z=\pi$.
Here, the system has 400 sites (N=400). The defect lies in the sites 201-204.}
\end{center}
\end{figure}

The Density of States (DOS) of these buckled line defects could present some
oscillations when compared to the silicene without any defect. This feature
occurs with both DFT and tight-binding calculations. This behavior is
illustrated in Fig. \ref{fig2},
where we present the DOS for the arm-2 defect. In this figure, we compare
the DOS calculated with a small supercell having 40 sites (same defect-defect
distance present in the ab initio calculations) with a DOS
obtained with a system 10 times bigger (N=400). In the latter system the line
defects
are separated by $48nm$ from its periodic image, and in fact there is no
interaction between them.
As a consequence, the wave function presented in Fig. \ref{fig2} (b) decays to
zero before
having any overlapping with the wave function of the neighbour line defect.
In this case the DOS is almost indistinguishable when compared to the pristine
case. On the other hand, for the
smaller supercell (N=40), the distance between the defect and its periodic image
is $3.8nm$. Thus, there is an overlap between the wave functions localized at
the defect and its periodic image, leading to the appearance of several small
spurious oscillations in the DOS. It is important to stress that this
interaction does not affect any of our main conclusions.


\begin{thebibliography}{99}
\bibitem{silicene_first} H. Sahaf, L. Masson, C. L\'eandri, B. Aufray, G. Le Lay and F. Ronci, Appl. Phys. Lett. {\bf 90}, 263110 (2007).
           

\bibitem{gr}
         K. S. Novoselov, A. K. Geim, S. V. Morozov, D. Jiang, 
         Y. Zhang, S. V. Dubonos, I. V. Grigorieva  and A. A. Firsov,
         Science {\bf 306}, 666 (2004).

\bibitem{exp1} B. Lalmi, H. Oughaddou, H. Enriquez, A. Kara, S. Vizzini, B. Ealet, and B. Aufray, 
                       Appl. Phys. Lett. {\bf 97}, 223109 (2010).

\bibitem{exp2} C. L. Lin, R. Arafune, K. Kawahara, N. Tsukahara, E. Minamitani, Y. Kim, N. Takagi, M. Kawai,
                       Appl. Phys. Express {\bf 5}, 045802 (2012).

\bibitem{exp3}  H. Jamgotchian, Y. Colignon, N. Hamzaoui, B. Ealet, J. Y. Hoarau, B. Aufray, J. P. Biberian,
                          J. Phys. Condens. Matter, {\bf 24},  172001 (2012).  

\bibitem{exp4} B. Feng, Z. Ding, S. Meng, Y.  Yao, X. He, P. Cheng, L. Chen, K. Wu,
                      Nano Lett. {\bf 12}, 3507 (2012).

\bibitem{exp5} A. Fleurence, R. Friedlein, T. Ozaki, H. Kawai, Y. Wang, Y. Yamada-Takamura, 
                            Phys. Rev. Lett. {\bf 108},  245501 (2012). 

\bibitem{exp6} P. Vogt, P. De Padova, C. Quaresima, J. Avila, E. Frantzeskakis, M. C. Asensio, A. Resta, B. Ealet and G. Le Lay 
               Phys. Rev. Lett. {\bf 108}, 155501 (2012).

\bibitem{silicene_first_dft1} K. Takeda and K. Shiraishi, Phys. Rev. B {\bf 50}, 14916 (1994);

\bibitem{silicene_first_dft2} G. G. Guzm\'an-Verri and L. C. Lew Yan Voon, Phys. Rev. B 76, 075131 (2007).


\bibitem{review_on_silicene} A. Kara, H. Enriquez, A. P. Seitsonen, L. C. L. Y. Voon, S. Vizzini, B. Aufray, H. Oughaddou,
                             Surf. Science Rep. {\bf 67}, 1 (2012).

\bibitem{so} C. C. Liu, W. Feng, and Y. Yao, Phys. Rev. Lett. {\bf 107}, 076802 (2011).
C. C. Liu, H. Jiang and Y. Yao, Phys. Rev. B {\bf 84} 195430 (2011).

\bibitem{efield1} Z. Ni , Q. Liu , K. Tang , J. Zheng, J. Zhou, R. Qin ,Z. Gao, D. Yu, and J. Lu, Nano Lett. {\bf 12} 113 (2012). 

\bibitem{efield2} N. D. Drummond, V. Z\'olyomi, and V. I. Fal'ko, Phys. Rev. B {\bf 85}  075423 (2012).

\bibitem{efield3} M. Ezawa, New J. Phys. {\bf 14}  033003 (2012).

\bibitem{add-metals} X. Lin and J. Ni, Phys. Rev. B {\bf 86}, 075440 (2012).

\bibitem{siesta}
         E. Artacho, D. S\'anchez-Portal, P. Ordej\'on, A. Garc\'ia  and J. M. Soler,
         Phys. Status Solid B {\bf 215}, 809, (1999).

\bibitem{lda}
         J. P. Perdew  and A. Zunger,
         Phys. Rev. B {\bf 23}, 5048, (1981).

\bibitem{theo1} S. Cahangirov, M. Topsakal, E. Akt\"urk, H. Sahin and S. Ciraci, Phys. Rev. Lett {\bf 102}, 236804 (2009).

\bibitem{theo2} S. Wang, Phys. Chem. Chem. Phys. {\bf 13}, 11929 (2011).

\bibitem{chiral-com} We have also studied an intermediate chiral configuration
[see Fig. \ref{fig-chiral}]. Since this geometry has higher energy than both 
zigzag or armchair, as shown in table \ref{tab1}, we do not analyse it in detail.

\bibitem{hopp-com} We have also locally changed the value of the transfer integral at the defect without any appreciable changes
in the band structure.

\bibitem{tersoff.ramann} J. Tersoff and D. R. Hamann, Phys. Rev. B {\bf 31}, 805 (1985).

\bibitem{exp-buckled}  L. Meng, Y. L. Wang, L. Z. Zhang, S. X. Du, R. T. Wu, L. F. Li, Y. Zhang, G. Li, H. T. Zhou, 
                       W. A. Hofer, A. Werner H. J. Gao, Nano Lett. {\bf 13}, 685 (2013).

\bibitem{gb1} O. V. Yazyev and S. G. Louie, Nature Mat. {\bf 9}, 806 (2010).

\bibitem{gb2} O. V. Yazyev and S. G. Louie, Phys. Rev. B {\bf 81} 195420 (2010).

\bibitem{PRL110.085504} L. Chen, H. Li, B. Feng, Z. Ding, J. Qiu, P. Cheng, 
                        K. Wu, and S. Meng, Phys. Rev. Lett. {\bf 110}, 085504 (2013).

\bibitem{si.surface} F. Chiaravalloti, D. Riedel, G. Dujardin, H. P. Pinto and A. S. Foster, Phys. Rev. B {\bf 79}, 245431 (2009).

\bibitem{tm-matheus} M. P. Lima, A. J. R. da Silva and A. Fazzio, Phys. Rev. B {\bf 84}, 245411 (2011).

\bibitem{amft} M. Amft, S. Leb\`egue, O. Eriksson and N. V. Skorodumova, J. Phys.: Condens. Matter {\bf 23}, 395001 (2011).

\bibitem{miwa1} W. H. Brito, R. Kagimura and R. H. Miwa, {\bf 85}, 035404 (2012).

\bibitem{miwa2} W. H. Brito, R. Kagimura and R. H. Miwa, Appl. Phys. Lett. {\bf 98}, 213107 (2011).

\end{thebibliography}
\end{document}